# Self-adaptation of *Pseudomonas fluorescens* biofilms to hydrodynamic stress

**Josué Jara[1,†], Francisco Alarcón[2,3,†], Ajay K. Monnappa[4,†], José Ignacio Santos[5], Valentino Bianco[2], Pin Nie[6], Massimo Pica Ciamarra[6], Ángeles Canales[7], Luis Dinis[8], Iván López-Montero[4,8,\*], Chantal Valeriani[8,\*], and Belén Orgaz[1,\*]**

[1] Departamento de Farmacia Galénica y Tecnología Alimentaria, Universidad Complutense de Madrid, 28040 Madrid, Spain
[2] Departamento de Estructura de la Materia, Física Térmica y Electrónica, Universidad Complutense de Madrid, 28040 Madrid, Spain
[3] Departamento de Ingeniería Física, Universidad de Guanajuato, 37150 León, Mexico
[4] Instituto de Investigación Biomédica Hospital 12 de Octubre (imas12), 28041 Madrid, Spain
[5] SGIker-UPV/EHU; Centro "Joxe Mari Korta"; Tolosa Hiribidea, 72, E-20018, Donostia - San Sebastián, Spain
[6] Nanyang Technological University, Singapore
[7] Departamento de Química Orgánica, Universidad Complutense de Madrid, 28040 Madrid, Spain
[8] Departamento de Química Física, Universidad Complutense de Madrid, 28040 Madrid, Spain.
**\* Correspondence:**
[†] These authors contributed equally to this work
\* Corresponding authors
Email: ivanlopez@quim.ucm.es, cvaleriane@ucm.es, belen@vet.ucm.es

## Abstract

In some conditions, bacteria self-organise into biofilms, supracellular structures made of a self-produced embedding matrix, mainly composed on polysaccharides, DNA, proteins and lipids. It is known that bacteria change their colony/matrix ratio in the presence of external stimuli such as hydrodynamic stress. However, little is still known about the molecular mechanisms driving this self-adaptation. In this work, we monitor structural features of *Pseudomonas fluorescens* biofilms grown with and without hydrodynamic stress. Our measurements show that the hydrodynamic stress concomitantly increases the cell density population and the matrix production. At short growth timescales, the matrix mediates a weak cell-cell attractive interaction due to the depletion forces originated by the polymer constituents. Using a population dynamics model, we conclude that hydrodynamic stress causes a faster diffusion of nutrients and a higher incorporation of planktonic bacteria to the already formed microcolonies. This results in the formation of more mechanically stable biofilms due to an increase of the number of crosslinks, as shown by computer simulations. The mechanical stability also lies on a change in the chemical compositions of the matrix, which becomes enriched in carbohydrates, known to display adhering properties. Overall, we demonstrate that bacteria are capable of self-adapting to hostile hydrodynamic stress by tailoring the biofilm chemical composition, thus affecting both the mesoscale structure of the matrix and its viscoelastic properties that ultimately regulate the bacteria-polymer interactions.

# 1 Introduction

Biofilms are microbial communities associated to interfaces in which cells are embedded within self-produced extracellular polymeric substances (EPS) or matrix (Costerton et al., 1999). The presence of biofilms in clinical settings and food facilities is a major issue, as the dwelling cells are far more resistant than their planktonic counterparts (Donlan, 2001; Fagerlund et al. 2017). Although several mechanisms have been described as responsible for biofilm recalcitrance (Lewis, 2007; Hoiby et al., 2010), the presence of the matrix is by far the most important. The major components of this matrix, in addition to water, are polysaccharides, proteins and extracellular DNA (eDNA) (Mann et al., 2012). Matrix quality and quantity depends on a plethora of factors such as nutrient availability, bacterial species integrating the biofilms and hydrodynamic factors (Flemming et al., 2010; Mann et al., 2012; Gloag et al., 2013; Pearce et al., 2019). Namely, the biofilm matrix is tailored depending on its surroundings. Such plasticity helps biofilms to adapt to environmental changes and to survive under rather harsh conditions (Stoodley et al., 2002; Hou et al., 2018). Currently, most of the strategies intended for biofilm removal have proven not to be entirely adequate, as they are often just focused on cells but disregard the mechanical part of these structures (Jones et al., 2011; Persat et al., 2015).

Recent studies suggest that the biophysical properties of the biofilms influence their mechanical behaviour promoting microbial persistence in certain niches. For instance, Gloag et al. (2018) measured the mechanical properties of two *Pseudomonas aeruginosa* phenotypic variants, mucoid and rugose small-colony variants (RSCV), isolated from the lungs of cystic fibrosis patients. They found RSCV colony-biofilms had a gradual progression to more elastic-solid behaviour and the EPS of the mucoid variant becomes stickier, suggesting a more stable EPS over time. Taken together, this could be a mechanism for these phenotypic variants of *P. aeruginosa* to persist in the lungs. Similarly, the biophysical properties of dental biofilms influence their mechanical removal from surfaces and may have a significant impact on the successful of oral hygiene strategies (Fabbri et al., 2015).

Although several studies have focused on the biofilm's mechanical properties under different conditions (Towler et al., 2003; Acemel et al., 2018; Boudarel et al., 2018; Charlton et al., 2019; Jana et al., 2020), little is known about the role of the polymer matrix interactions on the mechanical stabilization of the biofilm and how they evolve along the biofilm's growth under hydrodynamic stress. From a physical point of view, biofilms can be regarded as "living gels" that exhibit viscoelastic properties, wherein cells are active particles dispersed in a passive matrix (Klapper et al., 2002; Wilking et al., 2011). Cells are actively growing and exchanging material with their environment, thus contributing to the gel strengthening and/or weakening depending on external conditions (Rupp et al., 2005; Tallawi et al., 2017). For instance, Allen et al. (2018) using atomic force microscopy demonstrated that *P. fluorescens* biofilms growing under high-nutrient environments were softer and more adhesive than those developed under low-nutrient conditions, suggesting external factors not only affect matrix quantity but quality. Recent studies have demonstrated that *Pseudomonas* is one of the genera most abundant among the dominant microbiota on food contact surfaces after cleaning and disinfection (Fagerlund et al., 2017; Maes et al., 2019), being *P. fluorescens* the species whose prevalence is higher in food processing plants (Stellato et al., 2017). Although this species is not pathogenic it can act as "helper" for others to persist in food facilities, mainly using *P. fluorescens* matrix as a shelter and/or as an anchoring surface (Puga et al., 2018). Therefore, the understanding of its biofilm mechanical properties under hydrodynamic stress is essential for the design of more effective strategies to remove and to control biofilms.

Still, some questions remain unanswered. Do cells response to environmental stresses resulting into a more stable biofilm? Is there any change in the matrix composition affecting the biofilm adaptation to mechanical stress? A multidisciplinary approach is required to better understand biofilm behaviour. The goal of our work is to characterize the growth parameters of *P. fluorescens* biofilms developed with and without hydrodynamic stress, correlate them with their mechanical stability and unveil the physico-chemical interactions and mechanisms that are responsible for such adaptation to external stimuli. Our data demonstrate that bacteria adapt both their chemical and structural composition of the biofilm to specific requirements issued from particular environmental stresses. This might guide the development of new strategies to control biofilm formation in clinical and food settings.

## 2    Materials and Methods

### 2.1. Bacteria and growth conditions

*Pseudomonas fluorescens* B52, originally isolated from raw milk (Richardson & Te Whaiti, 1978), was used as model microorganism. Overnight precultures and cultures were incubated at 20 ºC under continuous orbital shaking (80 rpm) in 10 mL TSB tubes. Cells were then harvested by centrifugation at 4,000 g for 10 min, washed twice with sterile medium and their suspension $OD_{600}$ adjusted to 0.12. The bacterial suspensions were further diluted in order to start experiments at an initial concentration of $10^4$ cfu/mL.

### 2.2. Experimental system for biofilms development

Biofilms were developed on commercial 22x22 mm, thin microscope borosilicate glass-coverslips. These coverslips provide single-use, cheap, clean and undamaged smooth surfaces, without scratches or other microtopographic irregularities. Sixteen coverslips were held vertically by marginal insertion into the narrow radial slits of a Teflon carousel platform (6:6 cm diameter). The platform and its lid were assembled by an axial metallic rod for handling and placed into a 600 mL beaker (**Figure S1**). The whole system, *i.e.*, coverslips, carousel and the covered 600 mL beaker, were heat-sterilized as a unit before aseptically introducing 60 ml of the inoculated culture medium. Carousels were inoculated with the *P. fluorescens* suspension to start with an initial concentration of $10^4$ cfu/mL. To check the effect of shaking on biofilm formation, incubation was carried out at 20 ºC over 96 h both in a rotating shaker at 80 rpm and statically. Under these conditions, biofilm growth covered approximately 70% of the coverslips surface.

### 2.3. Cell recovery and counting

For viable cell retrieval and count, coverslips were aseptically withdrawn, and immersed into sterile 0.9% NaCl to detach weakly adhered cells. The attached cells were removed from the coupons' surface by swabbing both sides of the coverslips. Cells were then transferred into test tubes with 1.5 mL of peptone water and glass beads that were vigorously mixed in a vortex stirrer to break up cell aggregates, decimally diluted in peptone water, and pour-plated on Tryptone Soya Agar (TSA, Oxoid). Counting was performed after 48 h incubation at 30ºC. Ten independent experiments were carried out and two coverslips diametrally opposed were taken from each carousel at a time.

## 2.4. Surface coverage (%) and biomass (OD) determination

In order to measure the percentage of surface colonized by biofilms, coverslips (extracted as previously described) were stained for 2 min with a Coomassie Blue (Brilliant Blue R, Sigma) solution in an acetic acid/methanol/water (1:2.5:6.5) mixture. This step was repeated twice. After drying, the coupons were scanned using a HP Scanjet 300. Images were first binarized and then processed with the free software ImageJ (Schneider et al., 2012). For biomass quantification (cells plus matrix), the stained coupons were afterwards immersed into 4 mL of the same solvent mixture and the whole biomass was detached with sterile cell scrapers. After full homogenisation of this suspension, optical density (OD) was measured in a spectrophotometer using a wavelength of 595 nm. Bare coupons were stained the same way and used as controls.

## 2.5. Statistical analysis

Ten independent experiments were performed and two coverslips were sampled each time (n=10). Data were analysed using Statgraphics Centurion software (Statistical Graphic Corporation, Rockville, Md., USA). ONE-way Analysis Of VAriance (ANOVA) was carried out to determine whether samples were significantly different at a 95.0% confidence level ($p < 0.05$).

## 2.6. Confocal laser scanning microscopy (CLSM)

To visualize *P. fluorescens* biofilms, coupons were extracted from the carousel, then rinsed with sterile 0.9% NaCl and stained with Syto 13 (D9542, Life Technologies), a cell permeable fluorescent probe that binds to DNA and with FilmTracer SYPRO$^R$ Ruby biofilm matrix stain (Invitrogen). Thus, for image analysis, green corresponds to *P. fluorescens* cells and red corresponds to the extracellular matrix. Z-stacks of representative 0.12 x 0.12 mm regions of the air-liquid interphase of the biofilm (**Figure 1**) were acquired using a Nikon Ti-E inverted microscope equipped with a Nikon C2 scanning confocal module and Nikon Plna Apo 100X NA 1.45 oil immersion objective. Three-dimensional projections (Maximum Intensity Projection, MIP) of every image were reconstructed from z-stacks using the IMARIS 8.1 software (Bitplane AG, Zurich, Switzerland). To calculate the biovolume ($\mu m^3$), the MeasurementPro module of the above mentioned software was used. For this, three images taken from three different coupons were segmented into two channels, green and red, that were analysed to estimate the volume occupied by *P. fluorescens* cells and matrix, respectively. Maximum height of these reconstructed images was also measured. Data were expressed as the average ± SD (n=3).

## 2.7. Oscillatory Shear Rheology

The viscoelastic response of biofilms (~ 1 mL) was determined under oscillatory shear strain (Discovery HR-2 rheometer, TA instruments) using a circular flat plate tool (40 mm diameter). Temperature (± 0.1 °C) was controlled with a Peltier element assisted by an external water thermostat. Shear measurements were performed at a 1 mm gap between the peltier element and the plate tool. To avoid any solvent evaporation during the experiment, both the sample and the plate tool were covered with a Solvent Trap (TA instruments). A sinusoidal strain $\gamma$ of amplitude $\gamma_0$ was applied to the biofilm at a frequency $\omega$: $\gamma(t) = \gamma_0 \sin(\omega t)$. The shear stress was monitored $\sigma(t) = G^* \gamma(t)$, where $G^*$ is the shear viscoelastic modulus $G^* = G' + iG''$, where $G'$ is the storage modulus and $G''$ is the loss modulus. The shear viscosity was calculated as $\eta = G''/\omega$.

## 2.8. Nuclear Magnetic Resonance

$^1$H-$^{13}$C CPMAS NMR spectra were recorded on a 400 MHz BRUKER system equipped with a 4 mm MASDVT TRIPLE Resonance HYX MAS probe with 150 mg of lyophilized biofilm material. Larmor frequencies were 400.17 MHz and 100.63 MHz for $^1$H and $^{13}$C nuclei, respectively. Chemical shifts were reported relative to the signals of $^{13}$C nuclei in glycin. Sample rotation frequency was 12 kHz and relaxation delay was 5 s. The number of scans were 11140. Polarization transfer was achieved with RAMP cross-polarization (ramp on the proton channel) with a contact time of 5 ms. High-power SPINAL 64 heteronuclear proton decoupling was applied during acquisition.

## 2.9. Population dynamics model

The mathematical modelling of cell population is done using coupled ordinary differential equations for the density of cells, consisting on a modification of the logistic model (Verthulst, 1838). The logistic equation or Verthulst's model describes a first phase of exponential growth and a later saturation, a sigmoid curve observed in many reproducing populations (Micha et al., 2007), also in our experiment. It was later rediscovered by Pearl & Reed (1920) and has been widely used as a fundamental growth model in ecology and for bacterial populations in particular (Baranyi & Roberts, 1994; Fujikawa & Morozumi, 2004). It cannot nevertheless account for the later decay stage in population numbers (the so-called death phase) observed in systems in closed batch, as is the case in our experiment. To account for this effect, in our model both the carrying capacity of the medium and the growth rate depend on the concentration of some limiting nutrient. Some previous models based on Verhulst's logistic model implicitly or explicitly use resource availability in their description (Smith, 1963; Di Toro, 1980) in different mathematical ways. In our case we have followed Monod's very general and simple rule of proportionality between the growth rate and (a limiting) nutrient concentration (Monod, 1949) and plugged this proportionality directly into the logistic differential equation. To close the system of equations, one has to provide an evolution equation for the dynamics of the nutrient (self-degradation and consumption by bacteria) which we based on ideas in (Marsden et al., 2014). We have used a similar rationale to include the effect of nutrient availability in the carrying capacity of the logistic model. These two ingredients (dependence of logistic model on nutrient, and consumption of nutrient by bacteria) couple both differential equations which satisfactorily show the three observed stages (exponential growth, saturation and decay) using a unique set of differential equations and parameters for all the stages.

Finally, we have further developed our model by using two equations (one for cell density and one for nutrient concentration) for the planktonic population and another two for the biofilm population. These two populations can also interact by attachment or detachment of bacteria and nutrient diffusion. These evolution equations have been solved using Runge-Kutta method of order 4. Model equations and parameters are described in the supporting information in detail.

## 2.10. Computer simulations: biofilm growth and mechanical properties

To numerically study the biofilm initial growth, we make use of a modified ad hoc run-and-tumble code (LAMMPS). Bacteria are spherocylinders interacting via a short-range repulsion/short-range attraction (Lennard Jones-like) potential, mimicking the effective attraction induced by the excreted polymers (EPS), where the attraction strength is set to $0.2\varepsilon$ (see Supplementary Information). An isolated bacterium (whose size corresponds to that of *P. aeruginosa*), moves performing a run and tumble, whose parameters have been calibrated to mimic the dynamics of early stage growth (Ludewig et al., 2009). Specifically, the diffusion coefficient is $D \sim 0.7 \mu m^2/s$, whereas the velocity of a bacterium is $v \sim 0.2 \mu m/s$. We model growth by making the bacterium's length increasing with time. Once it reaches twice its size, it duplicates. We set the duplication time to the experimental value reported for *P. fluorescens* (approximately 1 hour at 20ºC). To numerically study a biofilm under shear, we simulate it via dissipative particle dynamics (DPD) in LAMMPS. We prepare two different systems composed by bacteria, polymers and solvent whose volume ratio corresponds to the experimental one for static or shaken biofilms. Having equilibrated both systems, we gradually form crosslinks between different molecules, apply an external shear (Raos et al., 2006) and compute the stress-strain curve.

## 3 Results

### 3.1. *P. fluorescens* produces biofilms with a higher bacterial density and thicker matrix when grown under shaking conditions.

*P. fluorescens* biofilms were grown statically and under shaking conditions using the batch system described in the Materials and Methods section. After 24, 48, 72 and 96 hours, we harvested biofilm's samples and measured the attached cell density, the biomass (cell + matrix) and the percentage of covered surface. The maximum population density was reached after 48h under shaking and after 24h under static conditions (**Figure 1A**). In the latter case, the cell density did not seem to vary over time neither to reach the same maximum (at 48h) as in the shaking case. Although shaking seems to prevent early adhesion, once attached, cells grew more efficiently, being the population over 2 log higher than that of the biofilms statically grown ($p < 0.05$).

To assess the effect of the hydrodynamic conditions on the ability of *P. fluorescens* to colonize the coupons surface, the % of area covered by its biofilms was determined (**Figures 1B** and **1C**). In shaking conditions, cells spread faster all over the surface, including the air-liquid interface and the submerged area reaching its maximum between 48h-72h incubation (between 50-60%). Later on (at 96h), a detachment was observed in the submerged part of the coupons, whereas fragments of the biofilm in the air-liquid interface still remained (**Figure 1C**). This suggests that the matrix at the air-liquid interface is stickier than the rest. When biofilms were developed under static conditions, the colonization pattern of the surfaces was limited by the air-liquid interface height, not surpassing this zone as in the case of shaking biofilms, which explained the lower values observed (**Figures 1B** and **C**). The submerged area of the coupons reached its maximum after 48h incubation but it appeared slightly colonized (**Figure 1C**). Again, shaking seemed to have a positive effect on both, biofilm spreading over the surface and biofilm retention, especially at the end of the incubation time. It should be noted that in this type of biofilm development system, the air-liquid interfase is rapidly colonized by *P. fluorescens* cells, as oxygen concentration there is higher promoting cell migration to this area. This colonization is much more intense in the case of biofilms developed under shaking conditions as the movement not only improves oxygen and nutrient diffusion but also the probability of the *P. fluorescens* cells to reach the surface of the coupons. Biomass values ($OD_{595}$), that include cells plus matrix, showed significant differences depending on the hydrodynamic stress. Under shaking conditions (**Figure 1D**), the highest OD value was reached at 72h, whereas under static conditions, the maximum was reached during the first 24h incubation time but remained unchanged over 48h. Interestingly, the OD maximum was one half of the one grown under shaking conditions, suggesting shaking seems to boost not only cellular growing but also matrix production.

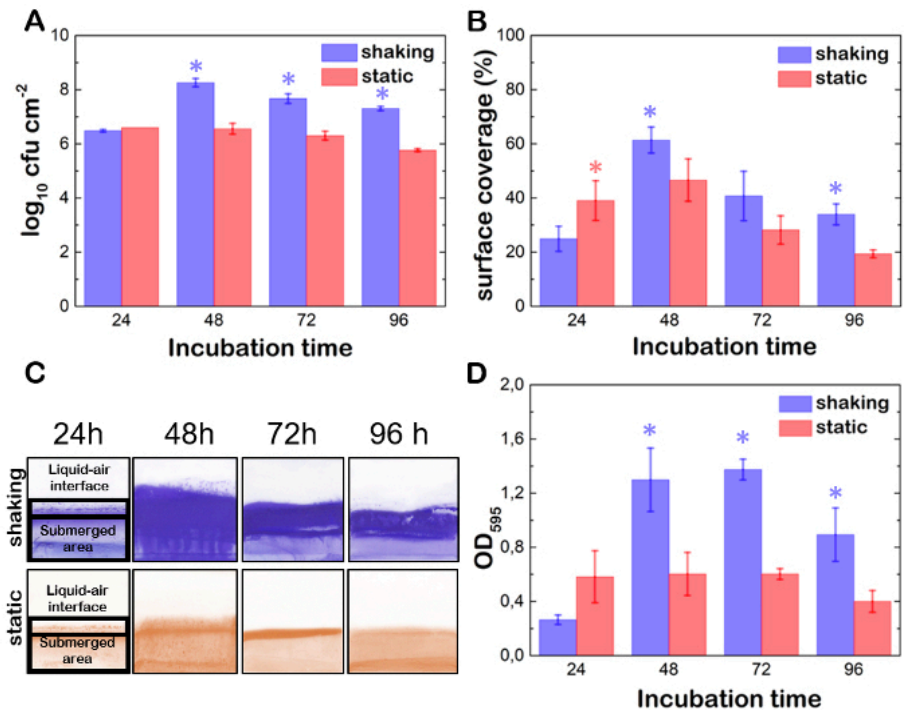

**Figure. 1. Evolution of *Pseudomonas fluorescens* B52 biofilms parameters under shaking (blue) and static conditions (red). A)** Attached cell population. **B)** % of surface coverage. **C)** Images of Coomassie blue stained coupons over time. **D)** Optical density proportional to the biomass production (OD) (cells+ matrix). Asterisks indicate statistically significant differences between shaking and static biofilms (n=10) ($p < 0.05$).

In order to better understand the relevance of shaking on the biofilm properties, Confocal Laser Scanning Microscopy (CLSM) images of *P. fluorescens* biofilms grown under static and shaking conditions were obtained. **Figure 2** displays the top views and the z profile of these biofilms. Under shaking conditions (**Figure 2**, **top panel**), the surface of the coupons was homogenously colonized. Moreover, biofilms produced a higher content of matrix (as shown in the corresponding pie chart). On the contrary, under static conditions (**Figure 2, down panel**), the colonization of the coupon surface was rather heterogeneous. CLSM images showed the presence of microcolonies scattered all over the surface, with empty areas in which dispersed cells were observed. In addition, less matrix was produced compared to the biofilms developed under shaking conditions (($1.4 \pm 0.1$) x $10^4$ µm³ vs. ($2.4 \pm 0.7$) x $10^4$ µm³, respectively). Indeed, the ratio matrix/cells was $0.9 \pm 0.1$ in the shaking case and $0.4 \pm 0.1$ in the static one (as shown in the corresponding pie chart). Moreover, shaking provided thicker biofilms with a maximum height of ($28 \pm 4$) µm vs. ($17 \pm 3$) µm in static biofilms (**Figure 2, z**

**profiles**), suggesting that the liquid movement favors the stacking of cell's layers and the matrix production as previously observed.

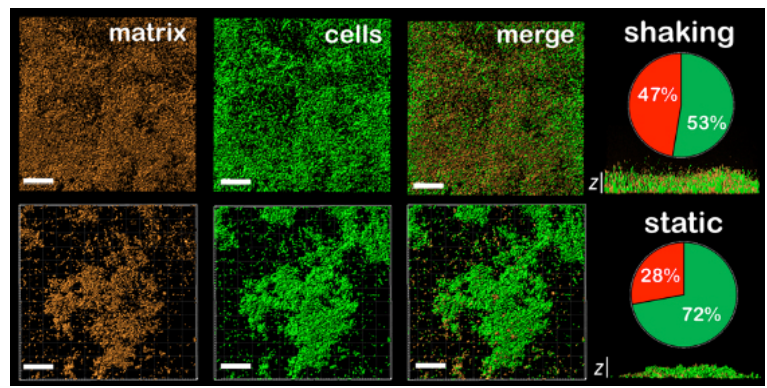

**Figure 2. CLSM images of different sections of 48h *P. fluorescens* biofilms developed at 20ºC under shaking (top panel) and static conditions (down panel).** The red channel corresponds to the zenital 3D view of *P. fluorescens* biofilm matrix stained with Sypro. The green channel corresponds to the zenital 3D view of *P. fluorescens* biofilm cells stained with Syto. Matrix appears in red and cells in green in merged images (scale bars are 20 µm). The *z* cross-sections show the width of a representative biofilm of those obtained, being the average of maximum height (28 ± 4) µm and (17 ± 3) µm in shaking and static biofilms, respectively (n=3) (scale bars are 20 µm). Pie charts represent the percentage of the volume occupied by cells (in green) and by matrix (in red) for each type of biofilm. Concretely, (53 ± 4) and (47 ± 4) for shaking biofilms and (72 ± 7) and (28 ±7) for static ones (n=3).

### 3.2. Shaking increases not only bacteria exchange rate from the planktonic to the biofilm but also the nutrient availability.

To test how shaking has an effect on biofilm cell counting, we developed a simplified model for biofilm formation based on a modified logistic population growth dynamics. The model is described by two variables: the bacterial population density in the biofilm ($\rho_b$) and the nutrient concentration (*c*) (*i.e.* carbon or nitrogen source, or oxygen). This concentration variable *c* effectively describes as a whole the nutrients needed for bacterial growth. The model's basic assumptions are: 1) the bacterial growth rate depends on the nutrient concentration (Monod, 1949). 2) The carrying capacity, the maximum sustainable population in the medium, depends on the nutrient availability (Smith et al., 2009). 3) Since experiments are performed in a closed batch, nutrients can only decrease in concentration (being utilized by bacteria or self-degrading) (see Supplementary Information). With these assumptions we recovered the experimentally observed exponential growth saturation-decay regimes.

Shaking or static conditions enormously affected the population at saturation point (*i.e.* the maximum population peak). One can expect that shaking may affect on the one hand nutrient diffusion and on the other hand bacterial transport (both from biofilm to plankton and vice versa). Regarding the nutrients, we focus on plankton-biofilm transport as we expect that the more scarce, more slowly diffusing nutrient from plankton to biofilm will control its growth. To include the effect of shaking, the model considers two evolving populations, both in the suspension (planktonic) and inside the biofilm, with their corresponding nutrient concentrations and with additional terms describing the rate at which bacteria attach/detach from the biofilm. Nutrients are depleted throughout the system, and diffuse proportionally to the concentration difference (see Supplementary Information).

A reasonable assumption is that shaking leads to an overall increase of transport, and results into an enhanced not only of the bacteria exchange rate from plankton to biofilm ($k^+$) but also of the nutrient diffusion coefficient (*D*). We cannot rule out however that the increased nutrient transport into the biofilm might also be favoured by the changes in matrix composition and structure we report in our NMR and rheology results. In addition, we consider that bacteria adhere more strongly to the surface to resist the hydrodynamic stress, and a lower detaching rate ($k^-$) was used in shaking conditions. This might indicate a stronger cell adhesion with the surface, which is consistent with a higher persistence of the shaken biofilm over time. Apart from this, the remaining parameters are set the same in both cases. As it appears clear from **figures 3A** and **3B**, the agreement with the corresponding experimental results is accurate. The differences in the biofilm growth rate under shaking and static conditions can be explained by an increase of transport efficiency of nutrients and bacteria. This is due to both a better mixing and a decrease of detachment rate, related to chemical changes in the matrix components leading to an effectively higher adhesion. Nonetheless, this logistic population growth dynamics model does not explicitly take into account the matrix production.

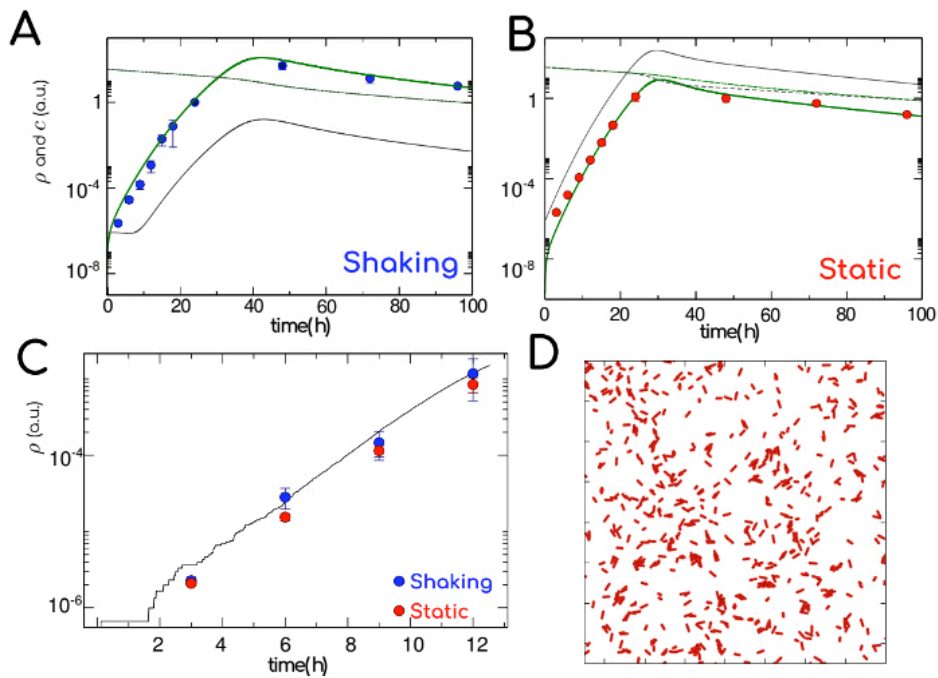

**Figure 3. Model for biofilm formation.** Population density in the biofilm (green thick line) and in the plankton (grey thin line), nutrient concentration in the biofilm (green dashed line) and in the plankton (grey dashed line). Experimental values (blue and red symbols, shaking and static respectively) normalized to the cell count at 24 h. **A)** Shaking: to mimic an efficient diffusion of nutrients and bacteria and strong adhesion respectively, we consider a high transition rate from plankton to biofilm (attachment) and diffusion ($k^+ = 1$ h$^{-1}$, $D = 1$ h$^{-1}$) and low transition rate from biofilm to plankton (detachment) ($k^- = 1 \text{ Å} \sim 10^{-3}$ h$^{-1}$). **B)** Static: to mimic a weak diffusion of nutrients/bacteria and poor adhesion, we consider a low attachment rate ($k^+ = 25 \text{ Å} \sim 10^{-3}$ h$^{-1}$) and diffusion ($D = 0.05$ h$^{-1}$) and high detachment rate ($k^- = 1$ h$^{-1}$) (low diffusion and low adhesion). **C)** Number of bacteria (log scale) versus time as calculated from full atomistic simulations (black line), reported together with the short time data in panels A and B. **D)** Onset of the biofilm formation, as in atomistic simulations. The elongated red particles mimic the bacteria, while polymers are simulated as an implicit effective attraction between bacteria since the first stages of bacterial growth happen on a surface, bacteria have been simulated in two dimensions. Data correspond to the average ± SD (n=10).

### 3.3. Effect of the polymers on the matrix structure during the biofilm formation.

To unravel the role played by the matrix in the early stages of biofilm formation, we perform *ad hoc* atomistic numerical simulations of run-and-tumble/self-replicating (active) elongated particles in two dimensions. Given that the presence of polymers translates into an effective depletion, active particles are considered as weakly attractive with each other. Simulating the bacterial dynamics in two dimensions (**Figure 3C** and Materials and Methods), we compute the bacteria growth rate as a function of time at very short times (corresponding to the experimental first 12h). For a weak attraction strength (see Materials and Methods) we recover the experimental short time behaviour, where shaking (**Figure 3C**, in blue) and static (**Figure 3C**, in red) results are indistinguishable. Given that the biofilm growth only happens in two dimensions when colonies consists of few hundred of cells (Beroz et al., 2018), we compute the number of cells (corresponding to very dilute concentrations): our atomistic results are in good agreement with our static and shaking experiments. Indeed, the density of bacterial cells as a function of time obtained from the numerical simulations (continuous black line) passed through the density of bacterial cells evaluated from shaking (red dots) and static (blue dots) experiments (**Figure 3C**). To summarise, within the first 12h of a biofilm formation, the bacterial growth rate is independent of the hydrodynamic stress and excreted polymers play the role of an effective (weak) attraction between replicating cells. We are left with the question on how the matrix production might alter the mechanical features of a biofilm.

### 3.4. The *P. fluorescens* biofilms are mechanically more stable under shaking conditions.

To study the mechanical behavior, a comprehensive shear rheological characterization was performed on harvested samples at different both incubation times and shaking conditions. Solid materials exhibit structural rigidity characterized by a finite shear modulus ($G' > 0$), whereas fluid materials are characterized by a zero shear modulus ($G' = 0$) and by a more or less high values of the loss modulus ($G'' > 0$), which defines a system able to flow under an applied force. These mechanical parameters can be obtained from the experimental stress-strain curves (see Materials and methods). In our case, the stress-strain curves are built by monitoring the stress response $\sigma(t)$ shown by the biofilm under a sinusoidal deformation $\gamma(t) = \gamma_0 sin(\omega t)$, where $\gamma_0$ is the amplitude of the deformation and $\omega$ the angular velocity of the deformation. In general, the stress-strain curves are mainly characterized by two stages. A first one, where the stress is proportional to the strain, that is, obeys the general Hooke's law, and the slope is the viscoelastic shear modulus. In this region, the material undergoes only elastic deformations. A second one, where the stress in no longer linear and the material present a plastic behavior. Both regimes cross at the yield point, which is characterized by a yield strain, $\gamma_Y$. Note that the shear viscosity of the material is related to the loss modulus by $\eta = G''\omega$ in oscillatory motion.

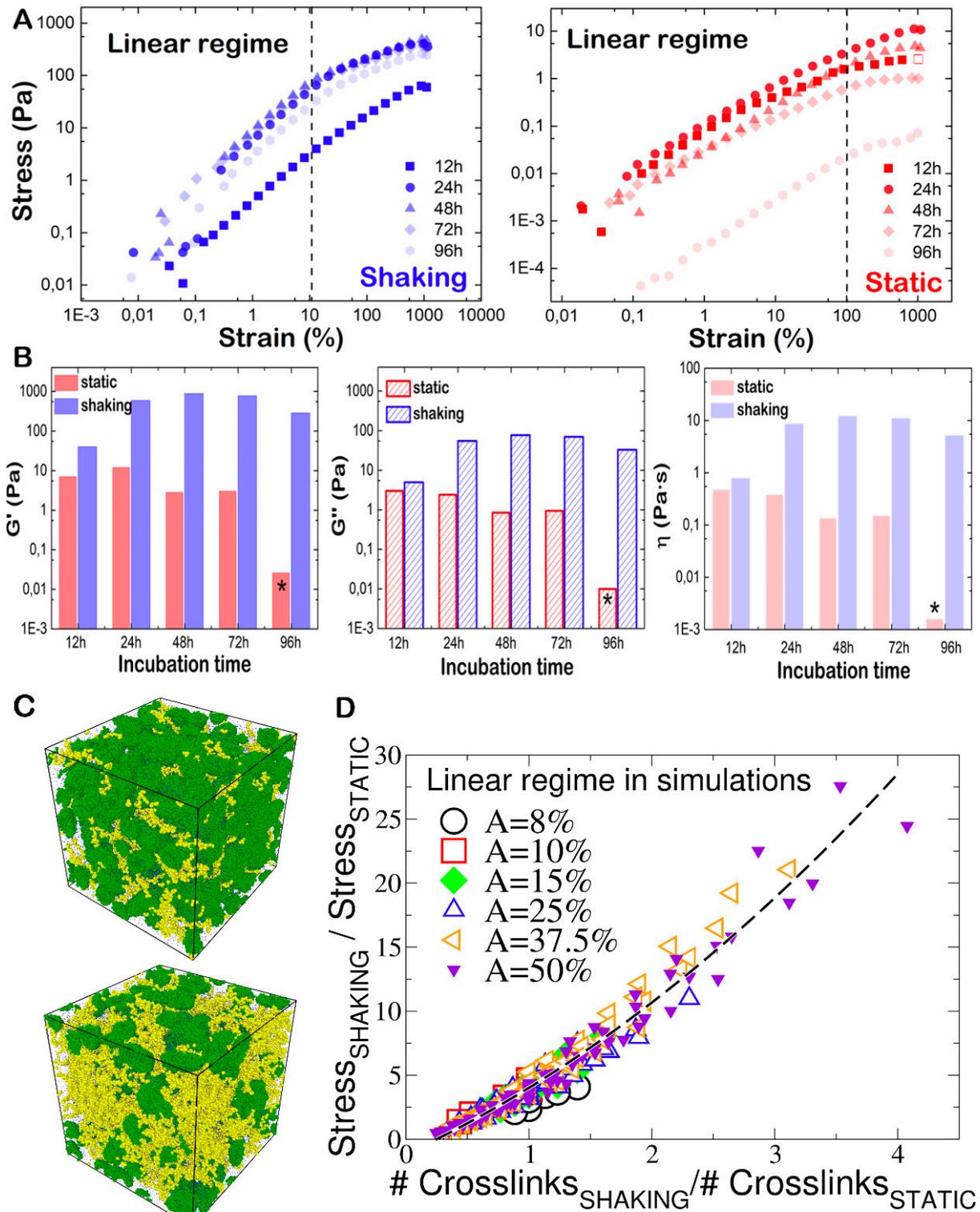

**Figure 4. Mechanical properties of *P. fluorescens* biofilms as mesured by rheology and computer simulations. A)** Stress–strain plot *P. fluorescens* at different incubation times under oscillatory shear ($f = 1$Hz and $T = 25$ºC). Biofilms were grown under shaking (left panel) and static (right panel) conditions. The mechanical response is found linear (dashed line) up to shear deformations of nearly ~ 20 and ~ 100% respectively, where a yield point is clearly visible in the $\sigma - \gamma$ plots. Beyond this limit the responses are nonlinear, and the two systems behave as a plastic body that yields under further stress. **B)** $G'$, $G''$ and $\eta$ values plotted as a function of the incubation time for *P. fluorescens* biofilms grown under shaking (blue) and static (red) conditions. Stars indicates that similar values were obtained for pure water. **C)** Snapshots of the simulated biofilm grown under static (up) and shaking (down) conditions. Green beads are bacteria, polymers in yellow. The beads size has been chosen for visualization. **D)** Ratio between the stress response of the biofilm grown under shaking and the one of the statically-grown biofilm, as a function of the ratio between the number of crosslinks in the two systems. Data computed with numerical simulations (Supplementary Information), and referring to different shear amplitudes within the linear regime response. The quadratic fitting function is $0.75x^2 + 4.4x - 1.14$ (black dashed curve).

**Figure 4A** shows the experimental stress-strain curves obtained for *P. fluorescens* biofilms under oscillatory shear at constant frequency of 1Hz. As expected, the stress responses in both systems are characterized by the elastic linear regime that reaches the plastic plateau above a yield strain $\gamma_Y \sim 0.2$ (20% strain) and $\gamma_Y \sim 1$ (100% strain) for shaking and static conditions, respectively. In either biofilm, we monitor a stress softening beyond $\gamma_Y$. To obtain the mechanical parameters G' and G'' in the linear regime, the rheological measurements were performed at a strain amplitude $\gamma = 1\%$, well below the yield point. The absolute values of G' and G'' are two and one order of magnitude higher, respectively, for shaken biofilms rather than for the static ones (**Figure 4B**). In other words, shaking leads to stiffer and more viscous biofilms, thus mechanically more stable and capable to resist the hydrodynamic environmental stress. In particular, the hydrodynamically stressed biofilm behaves as a pasty material with a relative high structural rigidity (G' > G'' > 0) rather than a viscous fluid (G' = 0 and G'' > 0). In the shaking case, after 24h, the values of shear viscoelastic moduli are G' $\sim$ 1kPa and G'' $\sim$ 100Pa. Although this parameter corresponds to a high viscosity as compared to a typical fluid, the fact that G' > G'' allows the system to be a quite resilient material with a relatively high mechanical compliance. In contrast, the shear moduli of static biofilms are similar in magnitude (G' $\sim$ G'' $\sim$ 10Pa). This translates in a very soft material, as shown by the relative high fluidity of the biofilm. Moreover, after 96h, the static biofilm displays an irreversible solid to fluid transition as no shear rigidity is monitored (G'=0). We guess that growth probably induces the static biofilm to lose its structural integrity as a soft material, which becomes a pure fluid (G' $\sim$ G'' $\sim$ 0). As derived from the G'' data, the shear viscosities of biofilms grown under shaking conditions are also one order of magnitude higher than viscosity of static biofilms over the whole range of biofilm growth (**Figure 4B**, right panel).

To better understand the different structure of either biofilm, we performed non-equilibrium molecular dynamic simulations of the biofilm grown in static and shaking conditions (**Figure 4C**), probing its viscoelastic response under an oscillatory shear stress of variable amplitudes, at 1 Hz frequency Within the chosen model, we control the number of crosslinks formed within the polymer-bacteria network, and compute the stress response. In the static case, the stress-strain curves (Supplementary Information) show a linear regime up to shear amplitudes of $\sim$ 100%, in perfect agreement with experiments (**Figure 4A, right**). As in experiments, for the shaken biofilm we detect a deviation from the linear regime at smaller amplitudes (**Figure 4A, left**), although the linearity holds up to $\sim$ 50%. We suggest that the discrepancy between numerical and experimental results is due to: i) the coarse-grain assumption that the biofilm topology is the same in both static- and shaken biofilm, contrary to experimental evidences; ii) the length scales of the biofilm components, involving two order of magnitudes (from $\sim$ μm size of bacteria to $\sim$ 100 nm of the polymers to $\sim$ 0.35 nm of the water molecules), inevitably reduced in the model. We compare the shaking/static stress response for the biofilm as function of the number of crosslinks in **Figure 4D**. Interestingly, all data collapse on a master curve, showing that the stress response is a growing function of the number of crosslinks, and allowing to estimate the increase of the shear response function between the two systems as function of the increase of formed crosslinks. Our simulations predict that the experimental ratio between the *G'* of the static and shaking biofilms (from **Figure 4B**, $G'_{shaking}/G'_{static} \sim 50$ at 24h and $\sim$ 150 at 48h) is recovered for a shaking biofilm containing from 6 times (at 24h) to 12 times (at 48h) the number of crosslinks present in the static biofilm.

## 3.5. Shaking increases the content of carbohydrates in the matrix of *P. fluorescens* biofilms.

To detect whether mechanical changes are correlated with changes in the matrix' composition, we studied biofilm samples via Solid-State NMR spectroscopy. This technique is especially suitable for studying insoluble biological materials. $^{13}$C CPMAS (Cross Polarization Magic Angle Spinning) solid-state spectra of intact biofilms obtained with and without shaking were measured for comparison (**Figure 5**). In both spectra we observed signals at 173 ppm from carbonyl groups of proteins, nucleic acids and phospholipids; aromatic signals at 120-160 ppm from aromatic amino acids and nucleic acids; signals at 95-105 ppm from anomeric carbons of carbohydrates; signals at 65-85 ppm from carbohydrate ring carbons; signals at 45-60 ppm from carbons at alpha position in amino acids and signals at 33-30 ppm from the $CH_2$ of lipids and amino acids. However, the relative intensity of the peaks in the spectrum of the shaken biofilm (**Figure 5A**) is different with respect to one of the spectrum of the static biofilm (**Figure 5B**). In general, the representative carbohydrates signals display a higher intensity in the shaken biofilm. This can be clearly seen when comparing the signal of the carbohydrate ring carbons (65-85 ppm) to that of the alpha carbons of amino acids (45-60 ppm), which displays similar intensity in both samples. The intensity of both signals are similar for shaken biofilms, differently from static biofilms. Similar observations are reported for the signals corresponding to the carbohydrate anomeric carbons. In addition, the biofilm grown under shaking shows a higher lipid content, since the signals at 173, 33 and 30 ppm displays higher intensity. Moreover, the increase in the intensity of the signals of the carbonyl carbons and the aromatic signals can be attributed to a higher proportion of nucleic acids on the biofilms grown under shaking conditions (both DNA and RNA could be compatible with the NMR signals detected).

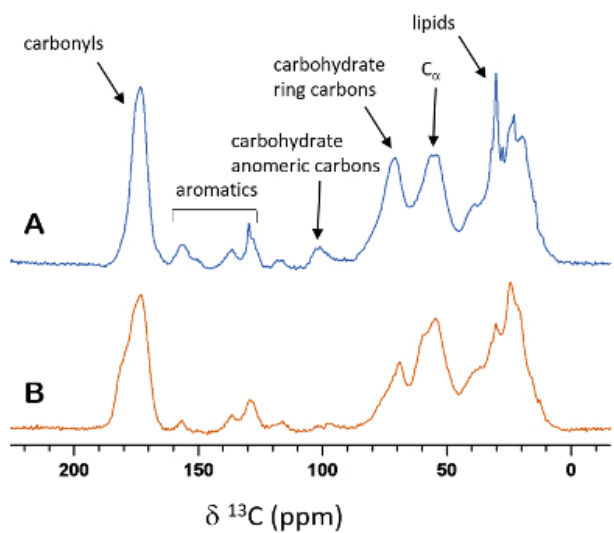

**Figure 5.** $^{13}$C CPMAS spectra of biofilm samples. **A)** Spectrum of a biofilm grown under shaking conditions. **B)** Spectrum of a biofilm grown under static conditions.

## 4 Discussion

Biofilms are considered forms of resistance for microorganisms to survive in hostile conditions. Indeed, antibiotic resistant clinical infections and persistent contamination in the food industry are often due to biofilms recalcitrance (Hoiby et al., 2015; ECDC, 2018; Alvarez-Ordoñez et al., 2019). Although not usually defined as virulence factors, biofilms decisively contribute to defend microorganisms, whether pathogenic or not, from any aggression, such as antimicrobial agents or the immune system (Otto, 2013). On the one hand, the biofilm itself is a microhabitat where the high population density together with the limiting diffusion due to the matrix, make the ecological relationships between cells quite complex. On the other hand, biofilms can be remodeled by environmental factors (Valderrama et al., 2013). The interplay between internal and external factors determines the characteristics of each particular biofilm, its stability and thus its bio-hazard level.

In our work, we have explored the effect of hydrodynamic stress on both growth and mechanical stability of *P. fluorescens* biofilms. Remarkably, all the typical features characterizing the biofilm growth and maturation were enhanced under shaking conditions. In particular, we monitored an increase in cell density, surface coverage and biomass (cell + matrix) production. Indeed, over time shaking biofilms OD values double those of static ones (on average 1.4 vs 0.6) (**Figure 1**). Obviously, shaking biofilms are denser in terms of cells (~8 log) than static ones (~7 log), but we think matrix production is somehow stimulated as these biofilms are thicker, stiffer and mechanically more stable over time. The structural basis for such stabilization could be sustained by both, a higher production of EPS and structural changes in matrix as suggested by rheological experiments and solid-state NMR (**Figures 4 and 5**). Hou and coworkers (2018) compared the EPS production of *Staphylococcus aureus* biofilms under different fluid shear conditions. They found that the amount of EPS was 5-times higher when a high shear (0.79 $s^{-1}$) was applied, suggesting changes in gene expression could be involved in such effect.

In our model we found that, when shaking, the concomitant effect of the biofilm growth and stability was physically favoured by a better nutrient accessibility through a diffusion-driven mechanism and by a faster bacterial exchange from the planktonic to the biofilm phase (**Figures 3A** and **3B**). The model thus incorporates two ways by which the biofilm population increases: the bacterial reproduction within the biofilm and the incorporation of bacteria from the planktonic phase to the already established biofilm. Although biofilms are rather dense structures, in a previous experiment we demonstrated that *Listeria monocytogenes* planktonic cells effectively penetrate *P. fluorescens* preformed biofilms (Puga et al., 2018). Also, Houry et al. (2012) demonstrated that several strains of bacilli were able to swim across *S. aureus* biofilms. Although in the present work we did not carried out such experiments, the assumptions underlying our model are grounded on previous works and on data here obtained, such as the population density and OD of both types of biofilms (**Figure 1**). Taking together, we assumed that over time planktonic cells can attach and penetrate a pre-existing biofilm. Additionally, shaking could favour the random movement of *P. fluorescens* planktonic cells increasing their probability to reach the coupon surface, as suggested by the wave-like pattern shown in Figure 1 at the interface of the biofilm. However, this mechanism would not work at short incubation times, where the growth rate was similar under both conditions (**Figure 3**). Nutrient availability and cellular exchange are not compromised at low cell density. In these conditions, the growth rate is governed by the bacterial duplication together with a weak attractive cell-cell interaction mediated by depletion forces issued by the presence of polymers (**Figure 3C**).

From a mechanically point of view, biofilms behaved as a composite material of hard inclusions (bacterial colonies) in a continuous soft matrix. The viscoelastic properties of this system are dominated by the compliant polymeric matrix as long as it prevails as the continuous phase (Arriaga et al., 2010). As both shaking and static biofilms can be considered as composite materials (see **Figure 2**), the different values of *G'* and *G''* measured by shear rheology (**Figure 4B**) have to be ultimately controlled by chemical differences in the matrix composition. As shown by solid-state NMR (**Figure 5**), *P. fluorescens* matrix displays a change in composition when developed under shaking conditions. Our current results point out an increase in the carbohydrate, lipid and DNA content in the biofilm samples grown with shaking. Polysaccharides of *P. fluorescens* biofilms have been described as mainly composed by glucuronic and guluronic acids, besides rhamnose, glucose and glucosamine (Kives et al., 2006), a polysaccharide very similar to those found in other species of *Pseudomonas*. For instance, *P. aeruginosa* is able to synthesize different types of carbohydrates, such as alginate, Psl (Polysaccharide synthesis locus) and Pel (Pellicle polysaccharide) (Mann et al., 2012) often associated with an increase in the adhesion properties of this microorganism (Chen et al., 2005; Harimawan et al., 2016). It has been recently demonstrated that Pel is a cationic polysaccharide that may interact with eDNA, reinforcing the biofilm matrix (Jennings et al., 2015). Indeed, mutations in the *Pel* locus make *P. aeruginosa* biofilms more susceptible to shear stress (Friedman et al., 2004). Although a further investigation is required to unequivocally assign the molecular origin of the biopolymers here revealed, the change in composition could be a key factor in the high stability of the *P. fluorescens* biofilms developed under hydrodynamic stress. In this line, Hartmann et al. (2019) have recently demonstrated how external fluid flows control the three-dimensional structure of *Vibrio cholera*e biofilms, suggesting biofilm cells could regulate the production of particular matrix components in order to modulate mechanical cell-cell interactions. In this context, we are working on the production of $^{13}$C labelled biofilms that will allow us to perform 2D solid state NMR experiments that are required to better characterize the observed changes.

The current strategies to fight against biofilms are mainly focused on two keystones: 1) killing the cells (Brooun et al., 2000; Hoiby, 2015; Chua et al., 2016) and 2) dispersing the matrix, mainly using enzymes (Kaplan, 2009; Johansen et al., 1997; Orgaz et al., 2006). Nevertheless, none of these have shown to be fully effective as they target both constituents separately leading in the worst case to the selection of resistant strains (Lewis, 2007; Vuotto et al., 2014; Chang et al., 2019; Yan & Bassler, 2019). Our results show that *P. fluorescens* is able to self-adapt to nonfavorable environmental stimuli increasing its biofilm stability. In particular, this microorganism changes the matrix chemical composition to completely fulfill two important requirements: a better interaction between the structural polymers and a stiffer scaffold able to resist the mechanical stress under shaking conditions. As a result, the biofilms become structurally more stable, increasing their inhabitants survival. Thus, knowing factors that increase the biofilm mechanical stability is a useful tool for predicting their danger and for guiding new strategies based on weakening the biophysical stability of these bacterial communities.

## 5  Conflict of Interest

The authors declare that the research was conducted in the absence of any commercial or financial relationships that could be construed as a potential conflict of interest.

## 6  Author Contributions

J, A.K.M., J.I.S. and A.C. performed experimental research, F.A., V.B., P.N., L.D. and M.P.C. numerical and theoretical research, and I.L-M., C.V and B.O. designed the research project and I.L-M., C.V., L.D., A.C., F.A., V.B. and B.O. wrote the manuscript.

# 7 Funding

This work was supported by the UCM/Santander grant PR26/16-10B (C.V., B.O. and I.L-M).

# 8 Acknowledgments

I.L-M, L.D and C.V. acknowledge financial support through grants PGC2018-097903-B-100, FIS2017-83706-R and FIS2016-78847. A.K.M. is recipient of a Sara Borrell fellowship (CD18/00206) financed by the Spanish Ministry of Health. F.A. acknowledges the support from a Juan de la Cierva fellowship and V.B. from the Marie Skłodowska-Curie Fellowship No. 748170 ProFrost.